\documentstyle[11pt]{article}

\begin{document}
\hbadness=10000
\setcounter{page}{0}
\title{%
\hspace*{\fill} {\normalsize DMR-THEP-93-8/W}\\*[-1.5ex]
\hspace*{\fill} {\normalsize hep-ph/9405232 }\\*[-1.5ex]
\hspace*{\fill} {\normalsize April 1994} \\[2ex]
{\huge \bf Photon interferometry of quark-gluon dynamics revisited}}
\author{
A. Timmermann\thanks{E. Mail: TIMMERMA@CONVEX.HRZ.UNI-MARBURG.DE} , M.
Pl\"umer\thanks{E. Mail: PLUEMER\_M@VAX.HRZ.UNI-MARBURG.DE} , L.
Razumov\thanks{E. Mail: RAZUMOV@CONVEX.HRZ.UNI-MARBURG.DE}{ } and R.M.
Weiner\thanks{E. Mail: WEINER@VAX.HRZ.UNI-MARBURG.DE} }

\date{Physics Department, Univ. of Marburg, Marburg, FRG }

\maketitle

\vspace{0.5cm}

\begin{abstract}
The Bose-Einstein correlations of photons emitted {}from a
longitudinally expanding system of excited matter produced in
ultrarelativistic heavy ion collision are studied.  Two effects found in
recent calculations -- that the correlation function in longitudinal
direction exhibits oscillations, and that it takes values below unity --
are demonstrated to be numerical artefacts and/or the results of
inappropriate approximations.  Thus, the general quantum statistical
bounds for the two particle correlation function of a chaotic source
with Gaussian fluctuations are confirmed. Two different expressions for
the two-photon inclusive distribution are considered.  Depending on
which of the two expressions is used in the calculation, the width of
the correlation function may vary by as much as $30\%$.\\[3ex] PACS
numbers: 05.30.Jp, 13.85.Hd, 25.75.+r, 12.40.Ee
\end{abstract}

\newpage

The study of high energy photons can yield information about the early
stages of the evolution of hot and dense matter created in
ultrarelativistic nuclear collisions. This is of special interest in the
experimental search for a quark-gluon-plasma (QGP), a new phase of
matter expected to exist for a brief period of a few fm/c before the
majority of the final state hadrons are emitted. In particular,
information on the space-time extension of the excited matter which can
be extracted {}from the two-photon correlation function may be used to
determine the energy densities reached in the collision.

Under the title ``Photon interferometry of quark-gluon dynamics'', a
calculation of the Bose-Einstein correlation (BEC) function of photons
emitted {}from a system of longitudinally expanding hot and dense matter
at RHIC and LHC energies was presented in Ref. \cite{sriv}.  The results
were later extended to include the effects of transverse flow, i.e., to
the case of a full three-dimenssional hydrodynamic expansion
\cite{sriv1,sriv2}.  It was argued\cite{sriv2} that the BEC's of photons
of high transverse momenta (on the order of a few $GeV/fm$) can be
sensitive to the presence of a mixed phase and may thus provide a
signature for the QGP.

It is the purpose of this note to point out that two rather surprising
features of the two-photon correlation function presented in Ref.
\cite{sriv} do not reflect genuine properties of the source. Rather,
they are artefacts of inappropriate approximations used to evaluate
space-time integrals in \cite{sriv}. To be specific, in Ref. \cite{sriv}
it was found that the BEC function in longitudinal direction (a)
exhibits oscillations and (b) takes values below unity. As was discussed
in Ref. \cite{bounds}, property (b) is inconsistent with the quantum
statistical bounds for a purely chaotic source. Below, it will be
demonstrated that both properties (a) and (b) disappear if the integrals
are calculated more exactly.

The BEC function of two identical particles can be written as
\begin{equation}
C_2(\vec{k}_1,\vec{k}_2) = \frac{P_2(\vec{k}_1,\vec{k}_2)}
{P_1(\vec{k}_1)P_1(\vec{k}_2)} \ ,
\end{equation}
where $\vec{k}_i$ ($i=1,2$) are the three-momenta of the particles, and
$P_2(\vec{k}_1,\vec{k}_2)$ and $P_1(\vec{k})$ are the one- and
two-particle inclusive spectra. For the general case of a Gaussian
density matrix and a purely chaotic source, the current formalism allows
to relate all $m$-particle inclusive distributions to a source function
$w(x,k)$ which describes the mean number of particles of four-momentum
$k$ emitted {}from a source element centered at the space-time point $x$
(cf. \cite{bounds} and refs. therein).  For the one- and two-particle
spectra, one has
\begin{equation}
P_1(\vec{k}) = \int d^4x \ w(x,k),
\label{eq:P1}
\end{equation}
and
\begin{equation}
P_2(\vec{k}_1,\vec{k}_2) = P_1(\vec{k}_1)P_2(\vec{k}_2) + \int d^4x_1
d^4x_2 \ w\left(x_1,\frac{k_1+k_2}{2}\right)
w\left(x_2,\frac{k_1+k_2}{2}\right) \cos(\Delta x^{\mu}\Delta k_{\mu}) .
\label{eq:P2}
\end{equation}
The expressions (\ref{eq:P1}) and (\ref{eq:P2}), which are usually
applied in calculations of the BEC function, have also been derived in a
Wigner function approach in \cite{pra}.  In Ref. \cite{sriv}, however, a
different form was used for the two-particle inclusive distribution,
namely,
\begin{equation}
P_2(\vec{k}_1,\vec{k}_2) = P_1(\vec{k}_1)P_2(\vec{k}_2) + \int d^4x_1
d^4x_2 \ w(x_1,k_1) \ w(x_2,k_2) \ \cos(\Delta x^{\mu}\Delta k_{\mu}) .
\label{eq:P2A}
\end{equation}
The form (\ref{eq:P2A}) was also derived in Refs. \cite{sin} as an
approximation to the complete, non-analytic result. As was shown in
\cite{bounds}, the expression (\ref{eq:P2A}) has the disadvantage that
in certain cases it can lead to results for BEC function which violate
the quantum statistical bounds for a purely chaotic source,
\begin{equation}
1 \ \leq \ C_2(\vec{k}_1,\vec{k}_2) \ \leq \ 2 \ ,
\label{eq:bou}
\end{equation}
whereas (\ref{eq:P2}) automatically satisfies Eq. (\ref{eq:bou}).

We have evaluated the above expressions (\ref{eq:P2}) and (\ref{eq:P2A})
for the case of thermal photons emitted {}from a longitudinally
expanding system of excited matter created in $Pb+Pb$ collisions at
$\sqrt{s}=200\ AGeV$. We use the source function \cite{sei} that was
adopted in \cite{sriv},
\begin{equation}
w(x,k)\ =\ const. \ T(x)^2 \
\ln\left(\frac{2.9k_\mu u^\mu(x)}{g^2 T(x)} +1\right) \
  \exp \left(-\frac{k_\mu u^\mu(x)}{T(x)}\right)
\end{equation}
where g is the QCD coupling constant, $T(x)$ is the local temperature
and $u^\mu(x)$ the local flow velocity.  To allow for comparison of our
results with those of Ref. \cite{sriv}, we used the same model for the
space-time evolution of the dense matter as in \cite{sriv}. That is to
say, we applied Bjorken hydrodynamics \cite{bjo} and adopted a massless
pion gas equation of state and a bag model equation of state for the
hadronic phase and the QGP phase, respectively.  The results presented
below correspond to an initial proper time $\tau_i=0.124\ fm/c$, an
initial temperature $T_i=532\ MeV$ and a freeze-out temperature
$T_f=140\ MeV$.

Fig. 1 shows the BEC function\footnote{Here and in the following, we
consider pairs of photons of equal polarization. Averaging over
polarizations leads to modifications which manifest themselves, among
other things, in a decrease of the intercept of the correlation
function.} in longitudinal direction, i.e., for a configuration of equal
transverse momenta, $\vec{k}_{1 \perp}=\vec{k}_{2 \perp}$. The
correlation functions are plotted against the rapidity difference
$\Delta y=y_1-y_2$.  The results presented in Fig. 1 refer to photons
emitted {}from the QGP phase only, at a transverse momentum
$k_{\perp}=5\ GeV/c$.

The curves labeled $1$ and $2$ were obtained by performing all
space-time integrations numerically. Curve $1$ corresponds to the
expression (\ref{eq:P2}) and curve $2$ to the expression (\ref{eq:P2A})
for $P_2(\vec{k}_1,\vec{k_2})$.  Neither of the two results shows
oscillatory behaviour, and they both respect the quantum statistical
bounds (\ref{eq:bou}). It is, however, interesting to note that the
width of the correlation function increases by about $30\%$ if
(\ref{eq:P2A}) rather than (\ref{eq:P2}) is used in the calculation. Let
us now consider the approximation adopted in Ref. \cite{sriv}. The
space-time integrals in (\ref{eq:P1}) -- (\ref{eq:P2A}) involve
integrations over the space-time rapidities $\eta_1$ and $\eta_2$. In
\cite{sriv}, the latter were performed analytically by using a Gaussian
approximation of the integrand as a function of $\eta_1$ and $\eta_2$ in
Eq. (\ref{eq:P2A}).  This procedure leads to the results displayed as
curve $3$ in Fig. 1.  The corresponding correlation function oscillates
as a function of $\Delta y$ and even takes values below one, i.e., it
violates the quantum statistical bounds (\ref{eq:bou}).  A comparison
with the ``exact'' results (curve $2$) shows that the Gaussian
approximation is inappropriate in this case. For completeness, we have
also included the cosine-approximation proposed in \cite{sriv} for small
$\Delta y$ (curve $4$ in Fig. 1),
\begin{equation}
C_2(\vec{k}_1, \vec{k}_2) \approx 1 + \cos[ 4 k_{\perp} \tau_i \sinh ^2
(
\Delta y /2)] .
\end{equation}

In Fig. 2 the BEC in longitudinal direction is shown for the sum of all
three contributions (hadron gas, mixed phase and QGP).  As was already
observed for the QGP component alone, at $k_\perp=5\ GeV$ the curves
obtained by using the expressions (\ref{eq:P2}) and (\ref{eq:P2A})
differ in width by a factor of $\sim1.3$. On the other hand, the
corresponding two curves for $k_\perp=2\ GeV$ are almost
indistinguishable.  It is noteworthy that while the $5\ GeV$ curves are
of of approximately Gaussian shape, the $2\ GeV$ curves are distorted
and suggest a two (or more) component structure. Fig. 3 compares the
plasma contributions to the sum of the contributions {}from all phases.
For $5\ GeV$ photons the plasma contribution dominates whereas for
transverse momenta of $2\ GeV$ the shapes of $C_2$ for the plasma
component and the sum of all components differ strongly. Indeed, as can
seen in Fig. 4 the transition {}from a two-component shape to a Gaussian
shape occurs in a narrow range of transverse momenta, $2\ GeV\leq
k_\perp \leq 3\ GeV$.  This behaviour can be understood in terms of the
relative importance of the contributions {}from the QGP, mixed and
hadronic phase. Fig. 5 shows the production rate of photons at $y=0$ as
a function of longitudinal proper time $\tau$ for five different values
of $k_\perp$. The two points in proper time where the slope changes
discontinuously correspond to the transition {}from the QGP to the mixed
phase and {}from the mixed phase to the hadron gas, repectively. For
high transverse momentum photons (e.g., $k_\perp=4\ GeV$), the QGP
contribution dominates, and consequently, there appears a single
Gaussian in the correlation function (cf. Fig. 3).  On the other hand,
for smaller transverse momenta ($k_\perp \sim 1\ GeV$) the contribution
{}from the mixed phase becomes comparable to or even exceeds the plasma
contribution, resulting in a two-component structure of $C_2$. Thus, if
such a change of shape as a function of $k_\perp$ will be observed it
may be considered as a piece of evidence for a first order phase
transition.\footnote{Of course, other effects can also give rise to a
two-component structure. One important example is quantum statistical
partial coherence \cite{rw}.}

In \cite{sriv2}, it was argued that a two-component form of the
two-photon correlation function in {\it transverse} direction may
signify the presence of a mixed phase and hence, of a first order phase
transition.  In Fig. 6 the BEC function is plotted against the component
$q_{out}$ of the transverse momentum difference (i.e., the component
parallel to the transverse momentum of the pair), for various values of
$k_{1 \perp}$. The results displayed in the figure which were obtained
by using the expression (\ref{eq:P2}) roughly agree with those obtained
in Ref. \cite{sriv} by using (\ref{eq:P2A}).  As in the case of the
correlation function in longitudinal direction, one observes the
deviation {}from the Gaussian shape as $k_\perp$ decreases, which
signifies the increasing importance of the contribution {}from the mixed
phase. Note that in contrast to the BEC function in $\Delta y$, the
correlation function in $q_{out}$ does show oscillations. However, it
does not take values below unity, i.e., it does not violate the quantum
statistical bounds (\ref{eq:bou}).

To summarize, we have demonstrated that two rather surprising properties
of the two photon correlation function presented in a recent publication
are artefacts of inappropriate approximations in the evaluation of
space-time integrals.  In \cite{sriv}, it was found that the BEC
function in longitudinal direction (a) oscillates and (b) takes values
below unity. As property (b) is inconsistent with general statistical
bounds, it is important to clarify the origin of this discrepancy. We
have shown that both properties (a) and (b) disappear if the space-time
integrations are performed numerically (i.e., without adopting any
analytic approximation).  On the other hand, we have confirmed that the
correlation function in transverse direction does exhibit oscillatory
behaviour in the {\it out}-component of the momentum difference, as was
observed in Ref.
\cite{sriv}. We have considered two different forms for the two-particle
inclusive distribution and found that the widths of the resultant
correlation functions may differ by up to $30\%$ depending on which one
of the two expressions is used in the calculation.  The change of the
BEC function in $\Delta y$ {}from a Gaussian to a two-component shape
with decreasing transverse momentum of the pair may serve as evidence
for the presence of a mixed phase, and hence, of a first order phase
transition.

Note that the results discussed above refer to a purely longitudinal
expansion. While there are indications\cite{udo} that at high collision
energies the expansion indeed becomes effectively one-dimensional, it
will be interesting to extend the present considerations to the case of
a fully three-dimensional hydrodynamic flow. However, this is a task
which exceeds the scope this brief report.\\[2ex]

This work was supported by the Federal Minister of Research and
Technology under contract 06MR731, the Deutsche Forschungsgemeinschaft
and the Gesellschaft f\"ur Schwer\-ionenforschung.

\newpage
\noindent
{\Large \bf Figure Captions}\\
\begin{description}
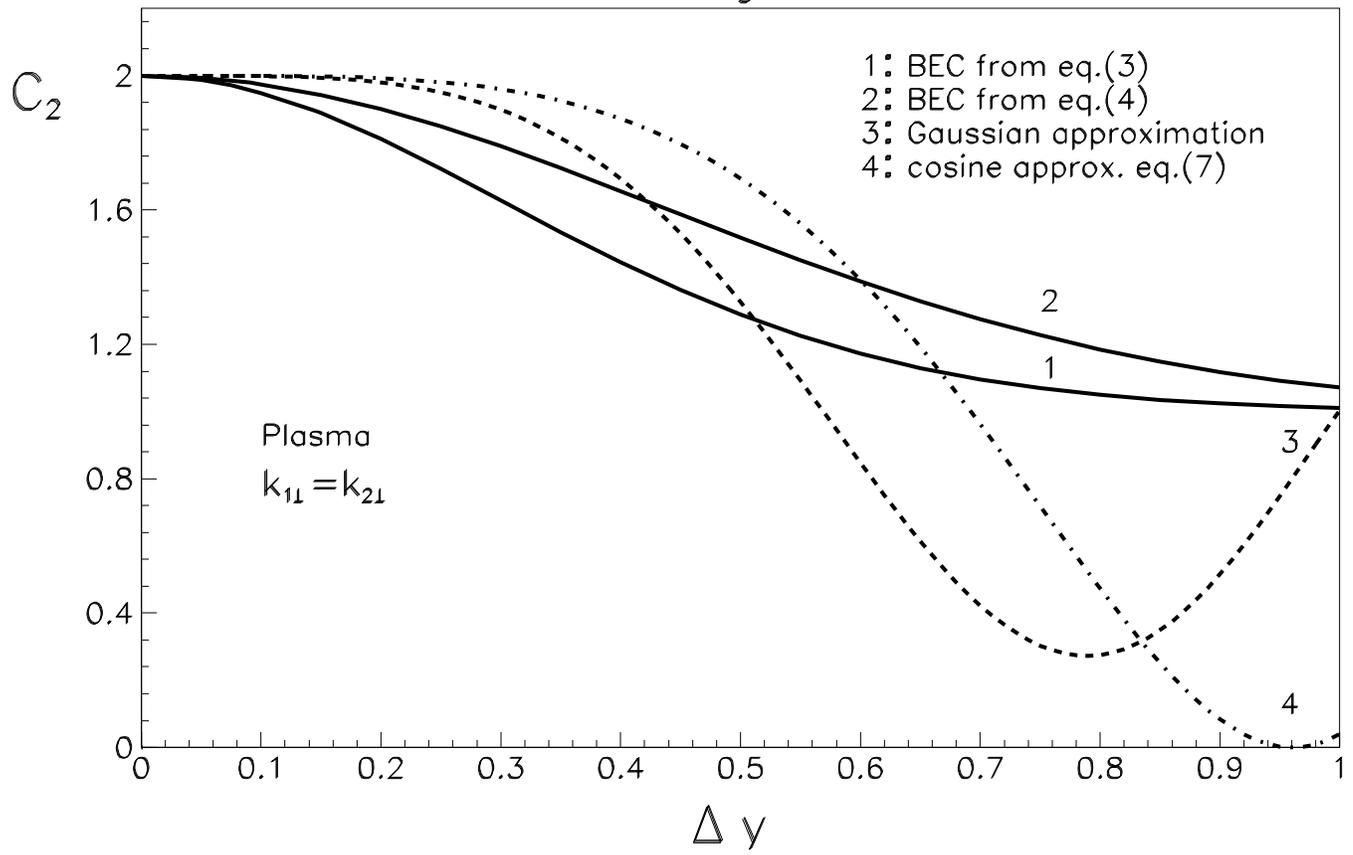
\item[Fig. 1] Two-photon Bose-Einstein correlation $C_2$
 as a function of the rapidity difference $\Delta y$, for photons of
transverse momentum $k_\perp=5\ GeV$ emitted in the QGP. The curves
correspond to different expressions and approximations for the
two-photon inclusive distribution.
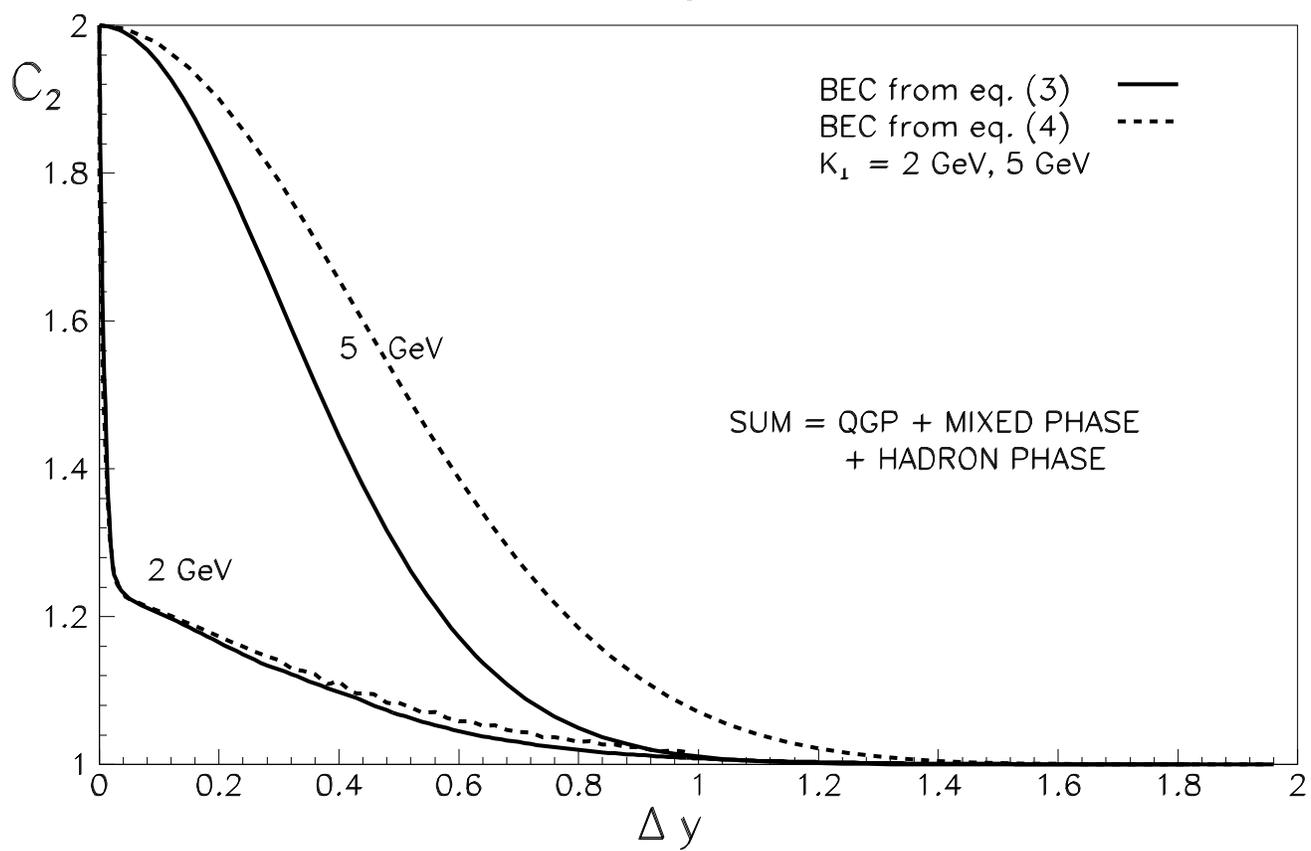
\item[Fig. 2] Comparison of $C_2(\Delta y)$  calculated
{}from the two expressions (\ref{eq:P2}) and (\ref{eq:P2A}) (see text).
The results include contributions {}from all three components (QGP,
mixed phase and hadron gas).
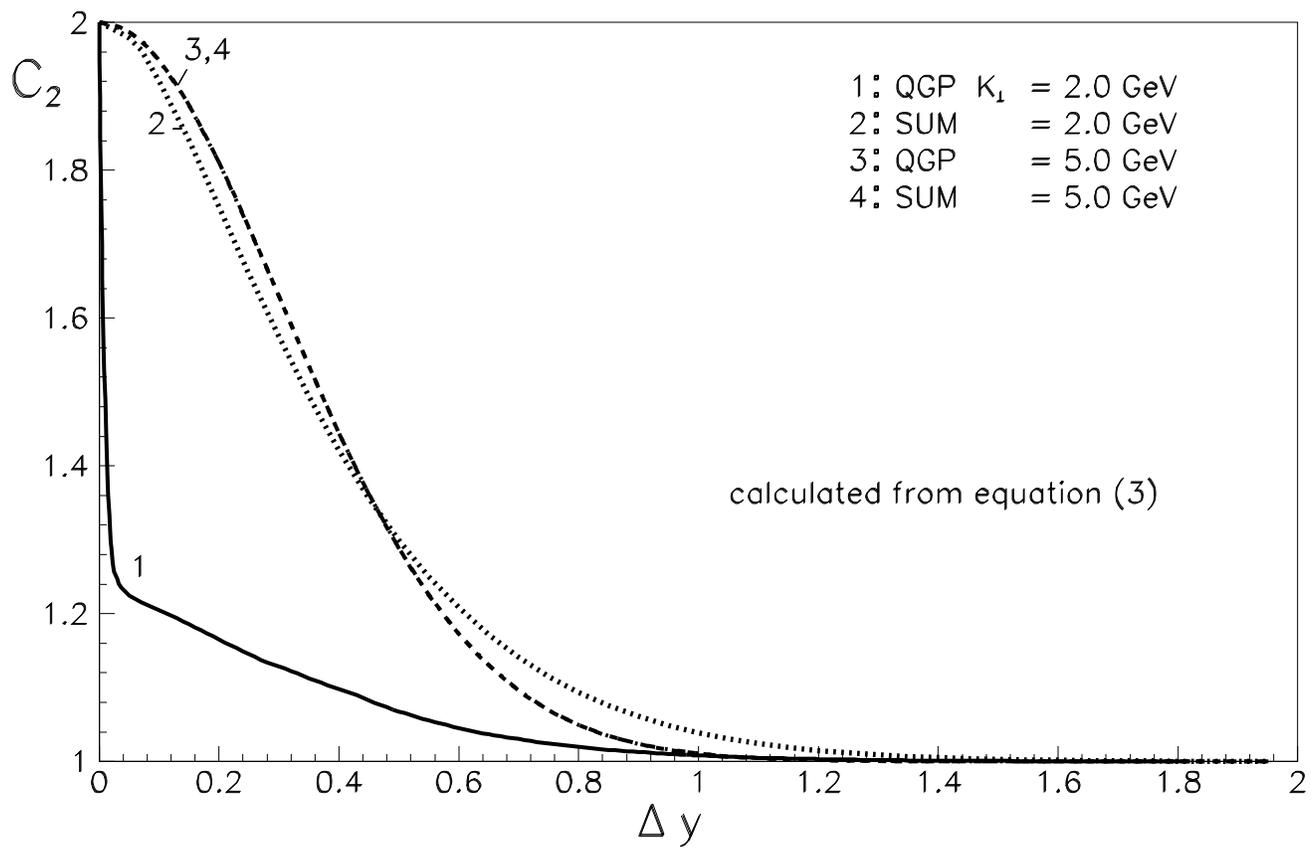
\item[Fig. 3] Comparison of the complete correlation function $C_2(\Delta y)$
and the separate contribution {}from the QGP phase.
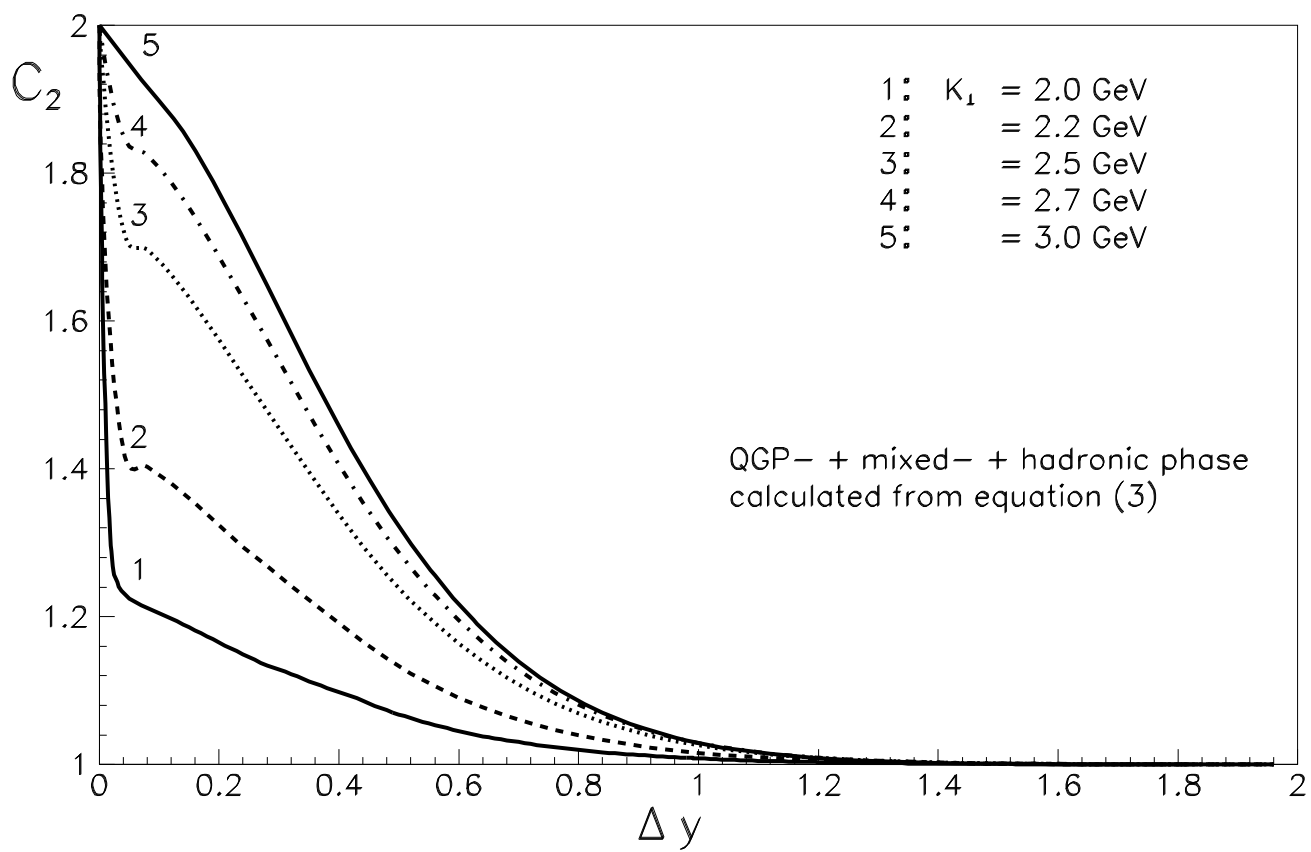
\item[Fig. 4] Dependence of the BEC function $C_2(\Delta y)$ on the
transverse momentum $k_\perp$ of the pair.
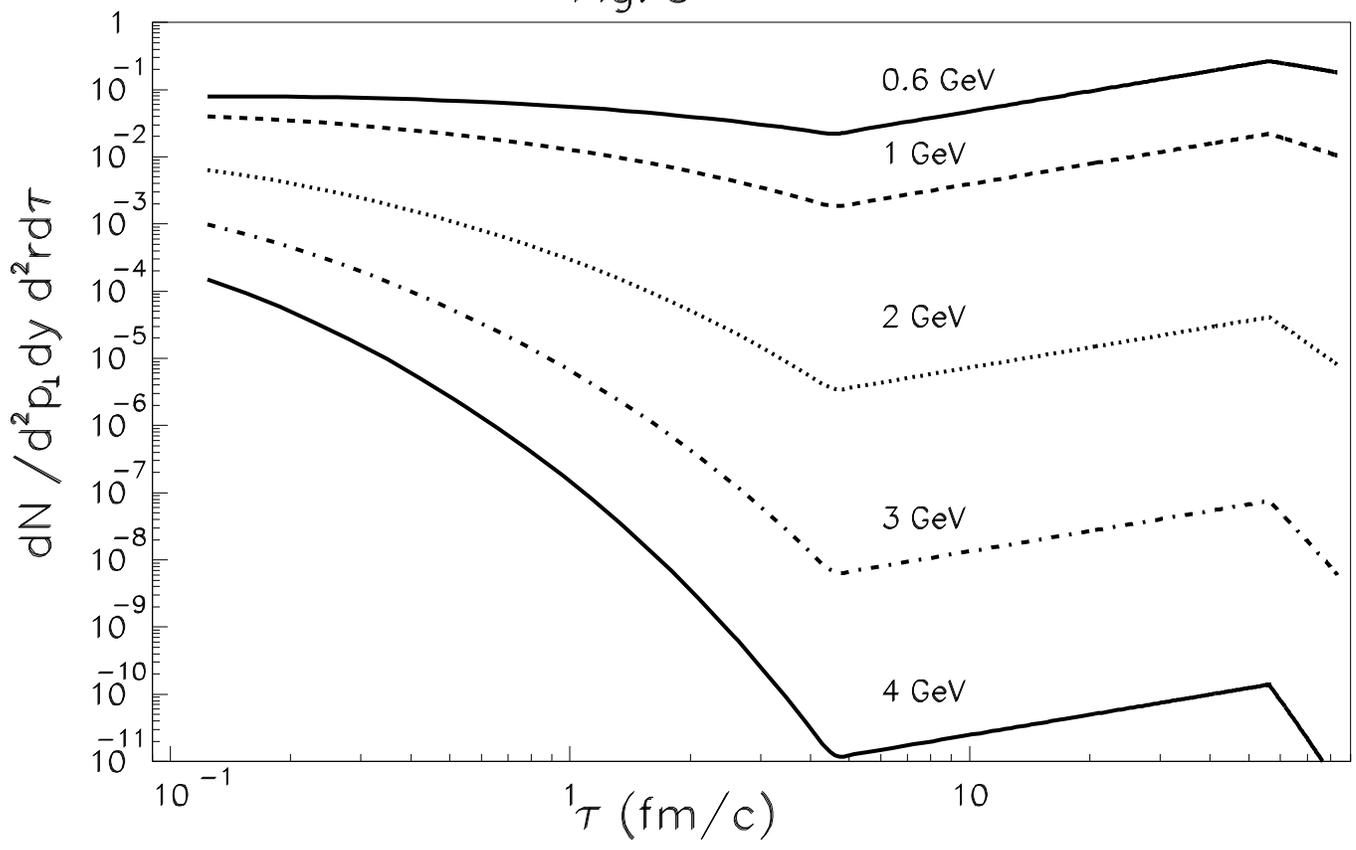
\item[Fig. 5] Dependence of the production rate of thermal photons on
the longitudinal proper time, for different values of the transverse
momentum.
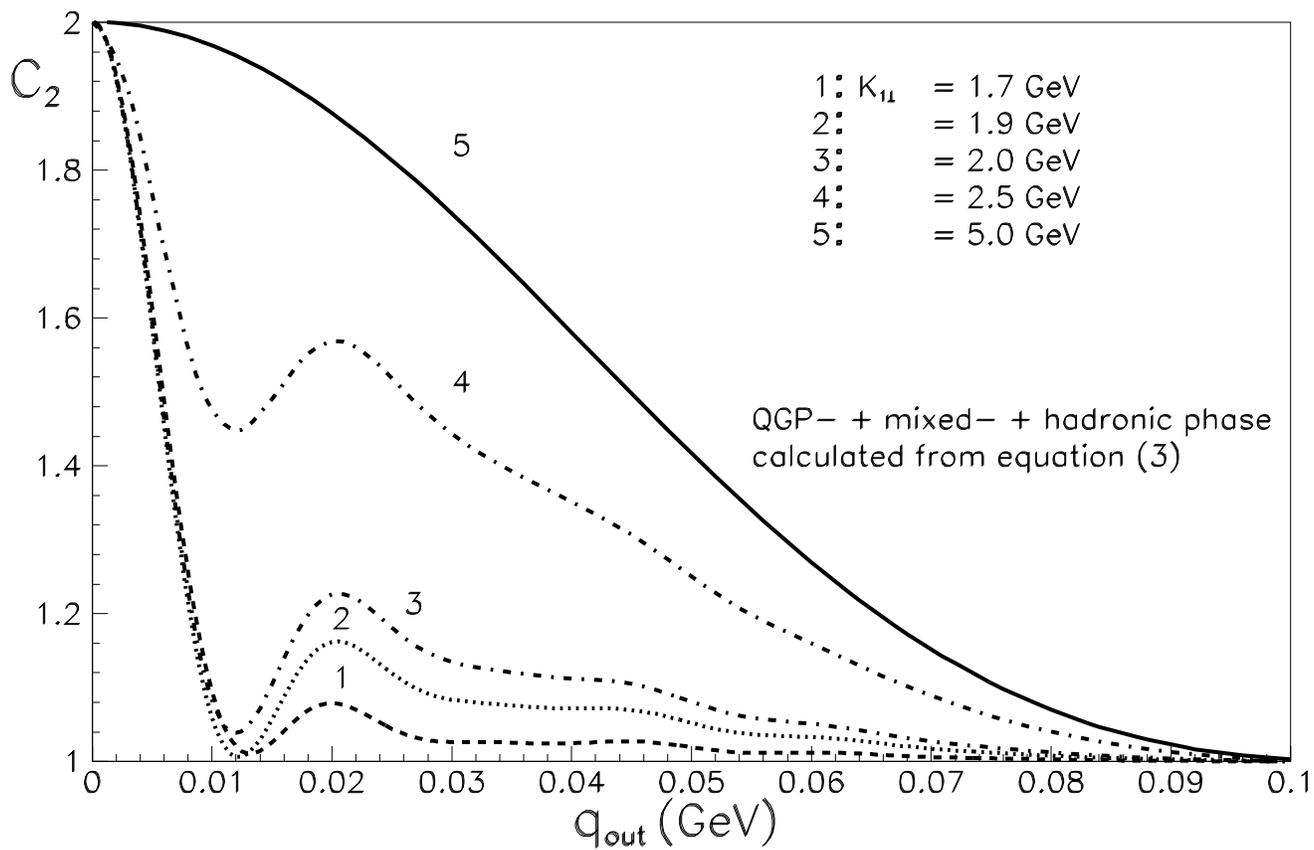
\item[Fig. 6] Two-photon BEC function in transverse direction,
as a function of the momentum difference in {\it out}-direction, for
different values of the transverse momentum $k_{1 \perp}$ of one of the
two photons.
\end{description}

\end{document}